%% file: main.tex
\documentclass[sigconf]{acmart}
\usepackage{pifont}
\usepackage{subfig}
\usepackage{cleveref}
\AtBeginDocument{%
  \providecommand\BibTeX{{%
    \normalfont B\kern-0.5em{\scshape i\kern-0.25em b}\kern-0.8em\TeX}}}

\begin{document}

\title{Transfer Learning for Security: Challenges and Future Directions}

\author{Adrian Shuai Li}
\email{li3944@purdue.edu}
\affiliation{%
  \institution{Purdue University}
  \streetaddress{}
  \city{West Lafayette}
  \state{IN}
  \country{USA}
  \postcode{}
}

\author{Arun Iyengar}
\email{ariyenga@cisco.com}
\affiliation{%
  \institution{Cisco Research}
  \streetaddress{}
  \city{San Jose}
  \state{CA}
  \country{USA}
  \postcode{}
}

\author{Ashish Kundu}
\email{ashkundu@cisco.com}
\affiliation{%
  \institution{Cisco Research}
  \streetaddress{}
  \city{San Jose}
  \state{CA}
  \country{USA}
  \postcode{}
}

\author{Elisa Bertino}
\email{bertino@purdue.edu}
\affiliation{%
  \institution{Purdue University}
  \streetaddress{}
  \city{West Lafayette}
  \state{IN}
  \country{USA}
  \postcode{}
}

\settopmatter{printacmref=false}
\setcopyright{none}
\renewcommand\footnotetextcopyrightpermission[1]{}
\pagestyle{plain}


\begin{abstract}

Many machine learning and data mining algorithms rely on the assumption that the training and testing data share the same feature space and distribution. However, this assumption may not always hold. For instance, there are situations where we need to classify data in one domain, but we only have sufficient training data available from a different domain. The latter data may follow a distinct distribution. In such cases, successfully transferring knowledge across domains can significantly improve learning performance and reduce the need for extensive data labeling efforts. Transfer learning (TL) has thus emerged as a promising framework to tackle this challenge, particularly in security-related tasks. This paper aims to review the current advancements in utilizing TL techniques for security. The paper includes a discussion of the existing research gaps in applying TL in the security domain, as well as exploring potential future research directions and issues that arise in the context of TL-assisted security solutions.

\end{abstract}
\maketitle

\input{sections/1_intro}
\input{sections/2_background}

\input{sections/3_applications}

\input{sections/4_challenges}
\input{sections/5_directions}
\input{sections/6_conclusion}

\balance

\bibliography{main}
\bibliographystyle{plainnat}
\end{document}

%% file: sections/1_intro.tex
\section{Introduction}

The cost of generating labeled data for a new learning task is often an obstacle to applying deep learning (DL) methods. A shift in data distribution at test time will likely degrade the performance of a trained DL model. One important example is in the context of automatic speed limit recognition from traffic signs, where one might have labeled data for traffic sign images from California while needing to predict speed limit from Indiana traffic signs. A model trained on the  California traffic signs will inevitably fail on the Indiana traffic signs. A promising approach to address such an issue is the use of transfer learning (TL) techniques by which knowledge, in the form of a pre-trained model or in the form of training data, can be transferred from one domain, referred to as {\em source domain}, to another domain referred to as {\em target domain} that has scarce training data. The appeal of transfer learning approaches is the ability to learn a highly accurate DL model that works well on the out-of-distribution target domain with only a few labeled target training data. 

In order to address the issue of limited data availability, conventional transfer learning (TL)-based approaches commonly employ a pre-trained model and fine-tune the trainable parameters using a small number of training samples from the target domain~\cite{sun2018deep,shao2018highly}. However, these pre-trained models are typically trained on large datasets like ImageNet~\cite{deng2009imagenet}, resulting in the inclusion of redundant features or irrelevant latent spaces that provide no significant benefits for the target inference tasks.

On the other hand, domain adaptation (DA)~\cite{ganin2016domain} aims to learn the target task by leveraging training samples from a source domain that is related to the target domain ~\cite{li2019diagnosing,wang2018deep, li2023maximal}. DA minimizes the discrepancy in latent space distributions between the source and target domains. This approach becomes particularly valuable when pre-trained models are not available or when the performance of pre-trained models after fine-tuning is unsatisfactory. However, it is worth noting that most existing DA approaches have primarily been evaluated on image classification tasks, with limited exploration on other types of data.

Several surveys have been conducted in the past few years on transfer learning and DA. For instance, Pan et al. categorized transfer learning into three sub-settings: inductive transfer learning, transductive transfer learning, and unsupervised transfer learning~\cite{pan2009survey}. On the other hand, Shao et al. classified transfer learning techniques into feature-representation-level knowledge transfer and classifier-level knowledge transfer~\cite{shao2014transfer}. Furthermore, all the previous surveys solely discussed 
DA in the context of image classification applications.

The problems of data scarcity (especially the attack data) and data drift are paramount, if not more important, in cybersecurity, making classification more difficult. 
transfer learning techniques offer promising solutions in the security domain, as they have the potential to enhance performance despite limited data availability and facilitate adaptability to emerging threats. In this paper, we focus on discussing transfer learning in security-related applications. Specifically, the key contributions of this survey are as follows: (1) we identify the security tasks that can be enhanced with transfer learning;   (2) we provide a detailed overview of the current research efforts that have successfully applied transfer learning in the security domain; (3) we analyze other security tasks to determine their suitability for leveraging transfer learning techniques; (4) we explore open problems in transfer learning for security and propose future research directions.

The remainder of this survey is structured as follows. In Section~\ref{background}, we first discuss common security tasks that can be improved using transfer learning, followed by an introduction of some notations and techniques in  transfer learning and motivations of using transfer learning in cybersecurity. In Section~\ref{methods}, we discuss past works on the application of transfer learning to security. In Section~\ref{problems}, we discuss the challenges and considerations in applying transfer learning for security. Finally, we propose the research directions and opportunities in Section~\ref{directions} and conclude the paper in Section~\ref{conclusion}.  

%% file: sections/2_background.tex
\section{Background}\label{background}

\subsection{ML-based security functions}
A comprehensive discussion of TL-based approaches for security is most effectively grounded in a taxonomy of ML-based security functions~\cite{bertino2023machine}. This taxonomy categorizes security techniques and processes for which machine learning methods have been applied. We now briefly discuss the categories of security functions that can gain advantages from employing transfer learning techniques.

\paragraph{Security policies learning}

Security systems such as access control systems, authentication systems, and network firewalls heavily rely on the implementation of robust and effective security policies. However, manually specifying these policies can be a time-consuming task and lacks scalability as the complexity of systems increases. To overcome these challenges, machine learning (ML) techniques ~\citep{abu2020polisma,abu2023flap} have been applied to automate the process of learning security policies, making it one of the early domains where machine learning has been successfully utilized.

Despite the significance of learning security policies through machine learning, it is worth mentioning that, to the best of our knowledge, there is a lack of recent work on transfer learning specifically in this domain. transfer learning techniques could potentially enhance the learning and adaptation of security policies by leveraging knowledge and models trained on related security tasks or domains.


\paragraph{Detection}
The detection of security-related events, such as intrusions, is a crucial aspect of ensuring effective security. As a result, numerous machine learning techniques have been proposed and developed to support intrusion detection systems over the years. In anomaly detection, the machine learning model learns the defined normal behavior and is then capable of distinguishing between normal and anomalous behavior (including 0-day attacks). Detection systems have been applied to different environment, such as networks~\citep{tang2020zerowall,mirsky2018kitsune}, cyber-physical systems~\citep{wang2018distributed, van2019real}, and IoT systems ~\citep{mudgerikar2019spion}. machine learning techniques make a significant contribution in improving the detection systems. This is also the area where transfer learning techniques have shown great success in dealing with scarce training data. We will discuss the state-of-the-art work in aiding network intrusion detection in Section \ref{nids}.

Another crucial aspect of detection pertains to malware detection. Machine learning (ML) approaches for identifying malware primarily rely on static features extracted from the malware binary. In Section \ref{malware}, we will delve deeper into how pre-trained vision classification models can assist in classifying malware that has been converted into images.

\paragraph{Software security analysis}
Software systems play a crucial role in various infrastructure and application domains. However, despite the awareness of the long-standing problem within the industry and research communities, software systems still remain vulnerable to security flaws. In light of this, there has been a recent focus on applying machine learning techniques for software security analysis. ML-based approaches in this context encompass a wide range of activities, from improving fuzzing techniques to achieve better coverage~\citep{rajpal2017not}, to enabling scalable static analysis for large code bases~\citep{lyu2022goshawk}. These initial approaches demonstrate the potential of machine learning techniques to enhance the effectiveness of software security analysis.  We expect this area will see many novel TL-based approaches to be developed. We will briefly discuss how transfer learning can further improve the function boundaries and assembly instruction recovery in Section \ref{XDA}.

\paragraph{Attack management}
Efficiently managing attacks is crucial to ensure the continued operation of a protected system. This involves taking defensive actions to detect attack stages.  block the next most probable attack and facilitate recovery. In Section \ref{HMM}, we will discuss how transfer learning can be combined with hidden Markov chains for detecting the attack stages in the network traffic, as well as, forecasting the next most probable attack stage. Additionally, conducting forensic activities is essential for identifying the vulnerabilities exploited during the attack. Although machine learning techniques have not been extensively utilized in attack management,  there is potential for their application in threat analysis, threat intelligence analysis and incident and response, particularly if relevant datasets were available. In Section \ref{forensic}, we will briefly discuss how the knowledge transfered from pre-trained LLM models can help with the downstreaming tasks such as forensic analysis.  



\subsection{Transfer Learning}

In what follows, we first give a formal definition of transfer learning, then introduce various categorizations of transfer learning based on domain shift and label availability. 

\paragraph{Notations and definitions}
We follow the definitions 
by Pan and Yang~\cite{pan2009survey}.  A domain $\mathcal{D}$ consists of a feature space $\mathcal{X}$ and a marginal probability distribution $P (X)$, where $X = \{x_1 , ..., x_n\} \in \mathcal{X}$. Given a specific domain $\mathcal{D = \{X} , P (X)\}$, a task $\mathcal{T}$ consists of a label space $\mathcal{Y}$ and an objective predictive function $f (\cdot)$, which can also be
viewed as a conditional probability distribution $ P (Y |X)$. In general, we can learn $P (Y |X)$ in a supervised manner from the labeled data $\{x_i, y_i\}$, where $x_i \in \mathcal{X}$ and $y_i \in \mathcal{Y}$.

Assume that we have two domains: the dataset with sufficient labeled data is the source domain $\mathcal{D}^s =
\{\mathcal{X}^s , P(X)^s \}$, and the dataset with a small amount of
labeled data is the target domain $\mathcal{D}^t =
\{\mathcal{X}^t , P(X)^t \}$.  Each domain has its own
task: the source task is $\mathcal{T}^s =\{\mathcal{Y}^s , P (Y^s |X ^s)\}$, and the target task is
$\mathcal{T}^t =\{\mathcal{Y}^t , P (Y^t |X ^t)\}$. In traditional deep learning, $P (Y^s |X ^s)$ can be learned
from the source labeled data $\{x^s_i , y^s_i \}$, while $P (Y^t |X ^t)$ can be learned from labeled target data $\{x^t_i , y^t_i \}$. 

\paragraph{Transfer Learning.}Given a source domain $\mathcal{D}^s$ and learning task $\mathcal{T}^s$, a target domain  $\mathcal{D}^t$ and learning task $\mathcal{T}^t$, transfer learning aims to help improve the learning of the
target predictive function $f_t (\cdot)$ in $\mathcal{D}^t$ using the knowledge in $\mathcal{D}^s$ and $\mathcal{T}^s$, where $\mathcal{D}^s \neq \mathcal{D}^t$ or $\mathcal{T}^s \neq \mathcal{T}^t$.

Based on the definition of transfer learning, the domain shift can be caused by domain divergence $\mathcal{D}^s \neq \mathcal{D}^t$ or task divergence $\mathcal{T}^s \neq \mathcal{T}^t$. Domain adaptation (DA) refers to the case where the source task $\mathcal{T}^s$ and the target task $\mathcal{T}^s$ are the same, and the domains are related but different.  

transfer learning techniques can be classified in two main categories based on distribution shift (homogeneous setting) or feature space difference (heterogeneous setting). In the homogeneous setting, the feature space between the source and target domains are the same with the same dimension. The distributions of the source and target data are different. On the other hand, in the heterogeneous setting, the feature space and the data distribution of the source and target domains are different. Regardless of which setting we have, we can further categorize the transfer learning techniques into supervised, semi-supervised and unsupervised settings based on the labeled data of the target domain. The last two settings are considered by most research work.

\subsection{Motivations of using TL in security}
\paragraph{Shortage of large scale high quality data}

The scarcity of large-scale, high-quality data presents a significant challenge in cybersecurity, hindering the creation of accurate machine learning models for identifying and managing threats. transfer learning emerges as a powerful solution to alleviate this issue in vision and natural language processing. By leveraging pre-existing knowledge from related domains or datasets, transfer learning enables transferring learned features or representations to the target domain, even when data is scarce or of lower quality. This approach allows models to capitalize on the knowledge encoded in larger, more diverse datasets from other domains and adapt it to the specific target tasks. Through transfer learning, cybersecurity practitioners can effectively enhance the performance of their models, despite the limitations posed by data scarcity, by tapping into the wealth of information available in other domains.

\paragraph{Improved performance with pre-trained models from other domains}

Applying machine learning (ML) in cybersecurity has garnered significant attention in prior research, yielding models surpassing conventional tools' efficacy while offering enhanced automation. Nevertheless, current ML-based methods still face complex challenges. Previous work has unveiled potential discrepancies in reported accuracies, attributing this degradation to inadvertent evaluations on testing datasets that substantially overlap with training data, thus impeding generalizability to real-world scenarios. Recent approaches have embraced a two-step transfer learning paradigm to address these concerns. The paradigm involves initial pretraining of the model to have a foundational understanding of the general task, followed by fine-tuning to cater to specific objectives. Notably, such transfer learning strategies have significantly improved model performance. For instance, in the XDA~\citep{pei2020xda} study, the model was pre-trained utilizing masked Language Modeling to establish a basic understanding of machine code. It subsequently refined its capabilities through fine-tuning for disassembly tasks. Leveraging semantic insights of machine code during fine-tuning has proven successful in accurately and reliably addressing tasks such as recovering function boundaries and assembly instructions.

\paragraph{Outdated models under emerging threats}
In real-world scenarios, threats continuously evolve over time. For example, new malware families are continuously evolving, which poses a great challenge in updating the models, especially when these data are limited. transfer learning offers a solution by facilitating the transfer of knowledge from old data to adapt the model to new instance. Rather than retraining the models from scratch each
time new threats are found, transfer learning seamlessly integrates newly collected data into existing models~\citep{singla2019overcoming,singla2020preparing}. This continuous model improvement ensures that the transfer learning model remains up-to-date
and easily adapts to the real world scenarios with continuously changing environment, leading to
enhanced performance over time. This ongoing improvement of the model ensures it can effectively adapt to the evolving real-world attack scenarios, enabling it to quickly identify new threats.

%% file: sections/3_applications.tex
\section{Applications of Transfer Learning in Cybersecurity}\label{methods}
In this section, we discuss prior research work involving transfer learning in cybersecurity. We discovered that transfer learning exhibits versatility across various security functions; however, its predominant application appears to be in intrusion detection and malware classification tasks. 

\subsection{Network intrusion detection}\label{nids}

A network intrusion detection system (NIDS) is a software or hardware system that identifies malicious network traffic usage by analyzing the patterns extracted from the packet capture and other network data source. NIDS are able to detect external attacks at an earlier phase, before the threats propagate to hosts and other networks. Because NIDS are a crucial building block for the security of network and computer systems, many techniques and tools are today available.

One of the well-known techniques for network attack detection is signature-based detection by which the system matches an intrusion signature with previously known signatures. However, signature-based detection approaches have difficulty in detecting zero-day attacks. A potential solution to this problem is to use anomaly-based detection approaches by which profiles on normal behaviors are generated. Then deviations from such profiles are flagged as anomalous. Anomaly detection techniques are often based on machine learning techniques. The use of machine learning improves  accuracy and reduces the need for human input.
Initial approaches for ML-based anomaly detection were based on support vector machines~\cite{li2012efficient}, decision trees~\cite{rutkowski2013decision}, and nearest neighbour methods~\cite{lin2015cann}.

However, those initial ML-based anomaly-based detection approaches suffered from high false positive rates. In recent years, DL has achieved significant success in anomaly-based intrusion detection. With one or more hidden layers, DL classification models are able to produce highly nonlinear models which learn the complex relationship that exist between the input data and the labels, namely intrusion or normal, and classify the unseen data at the testing stage.  However, DL methods work well only under the assumption that the training and test data are drawn from the same domain where they have the same feature space and same distribution. 

As new attacks discovered, the labeled data obtained in one time period may not follow the same distribution in a later time period, making the trained model work poorly and unable to detect the new attacks. Gathering again a large amount of labeled samples and rebuilding the model is expensive and time consuming. Further, one may not be able to directly apply a classifier learned from one network to another network when the two networks differ with respect to characteristics and/or traffic patterns.

To address the above problems, recent approached have leveraged transfer learning techniques, which also have been concerned with the problem of domain shift. In essence, the source and target domains represent different or same networks with different attacks captured at different times. We assume that the source domain has a large amount of labeled data and the target domain has a few or no labeled data. The source and the target datasets may have the same or different feature space. Further, the target may have new attack types not present in the source. The goal is to accurately detect both the new attack types as well as the old attack types. 

Singla et al.~\cite{singla2019overcoming} utilize pre-trained models from related tasks for NIDS with limited training data. Their approach involves initially training a deep neural network model on the source data, which is similar to the target data. Then, they employ transfer learning to fine-tune the model using the target data. The limitation is that this technique cannot be directly applied to scenarios where the source and target datasets have different feature space.

More recently, Zhao et al.~\cite{zhao2019transfer} have focused on learning features from a source domain that can be adapted to the target domain. They computed Euclidean-based similarities between the source and target domains and utilized these similarities as input for a classifier to identify unknown network attacks.

Another approach by Singla et al.~\cite{singla2020preparing} uses a DA-based transfer learning technique to identify evolving attacks with minimal new training data. Adversarial DA leverages a similar idea to generative adversarial networks (GANs)~\cite{goodfellow2020generative}.  Historically, the use of GANs focused on generating data from noise. Its main goal is to learn the data distribution and then create synthetic examples that have a similar distribution. 

A GAN consists of a generative model, called generator $G$, and a discriminative model, called discriminator $D$.  The generator $G$ generates data that are indistinguishable from the training data and the discriminator $D$ distinguishes whether a sample is from the data generated by $G$ or from the real data. The training of the GAN is modeled as a minimax game where $G$ and $D$ are trained simultaneously and get better at their respective goals: training $G$ to minimize the loss in Eq.~\ref{eq1} while training $D$ to maximize it: 
\begin{equation} \label{eq1}
    \min_{G}\max_{D} V(G,D) = E_{x}[logD(x)]  + E_{z}[log(1-D(G(z)))]
\end{equation}
where $E_x$ is the expected value over all real instances, $E_z$ is the expected value over all the generated data instances, $D(x)$ is the probability of $D$ predicting a real instance as real, and $D(G(z))$ is the probability of $D$ predicting a generated instance as real instance.  

\begin{figure}[t]
  \centering
  \includegraphics[width=\linewidth]{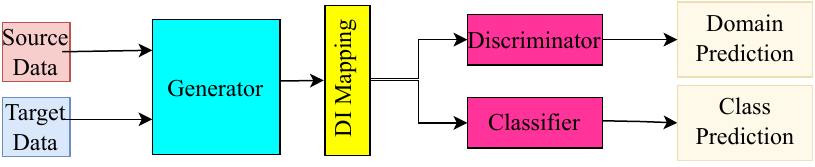}
  \caption{GAN architecture for adversarial DA ~\cite{singla2020preparing}. }
  \label{GAN}
\end{figure}

In adversarial DA~\cite{ganin2016domain}, this principle has been employed to ensure that the network cannot distinguish between the source and target domains by learning features that combine 
discriminativeness and domain invariance. 
The architecture is shown in Figure~\ref{GAN}. 
This is achieved by learning domain invariant features  as well as training two classifiers operating on these features: \ding{182} the label predictor that predicts class labels and \ding{183} the discriminator that discriminates between the source and the target domains during training. The discriminator is trained to minimize the loss in Equation \ref{eq2.2} and the label predictor is trained to minimize the loss in Equation \ref{eq2.3}. The shared generator updates its weight to minimize the loss of the label classifier (by minimizing Equation \ref{eq2.3}) and maximize the loss of the discriminator (by minimizing Equation \ref{eq2.4}). The generator works adversarially to the discriminator, encouraging domain invariant features to emerge in the course of optimization. 
\begin{align}
    L_d & = - E_{x_s}[log D(G(x_s))]  - E_{x_t}[log(1-D(G(x_t)))]\label{eq2.2}\\
    L_c & = - E_{x_{attack}}[log C(G(x))]  - E_{x_{benign}}[log(1-C(G(x)))]\label{eq2.3}\\ 
    L_g & = - E_{x_t}[logD(G(x_t))]\label{eq2.4}
\end{align} 
where $E_{x_s}$ and $E_{x_t}$ are expected values of the source and target samples, $D(G(x_s))$ is the probability of predicting a source domain sample as belonging to source, and $D(G(x_t))$ is the probability of predicting a target domain as belonging to source. $E_{x_{attack}}$ and $E_{x_{benign}}$ are the expected values of attack and benign samples, $C(G(x))$ is the probability of the classifier predicting a sample as attack and $1-C(G(x))$ is the probability of classifying a sample as benign.

\subsection{Malware detection and classification}\label{malware}

Another way transfer learning can be used is by leveraging models trained on large datasets. For example, previous works in computer vision have shown that using pre-trained models on ImageNet~\cite{deng2009imagenet}
can improve the accuracy of target task while reducing the amount of labeled data required for training.  The success in vision has driven the design of approaches that leverage pre-trained models on large image datasets to the downstreaming malware detection task.

Rustam et al.~\cite{rustam2023malware} focused on malware prediction using transfer learning. They developed a bimodal approach, which encompass the extraction of features using VGG and ResNet models. These extracted features are subsequently employed as inputs for machine learning models. In a different work, bytecode data was transformed into images and used as the input for an MCFT-CNN model~\cite{kumar2021mcft}. This model leverages low-level features from ImageNet to classify malware. Similarly, another method called DTMIC~\cite{kumar2022dtmic} converted binary data into images and used ImageNet for malware classification. 

\subsection{Software security analysis}\label{XDA}

Previous approaches have explored using machine learning for the disassembly of binaries. The resulting models outperform the accuracy of traditional disassemblers at recovering assembly instructions and function boundaries. However, these methods perform poorly when the testing data significantly shifts from the training data. Furthermore, recent research shows that these methods are not robust to compiler optimization changes. In \cite{pei2020xda}, a new ML-based disassembly framework called XDA is introduced that uses transfer learning to address these challenges. XDA first pre-trains the model using a masked language modeling task which asks the model to predict missing bytes given the machine code. This design forces the model to learn dependencies between masked and surrounding bytes. In the second stage, XDA fine-tunes the model to solve a specific disassembly task.

Transfer learning combined with masked language modeling  can solve popular disassembly tasks accurately. XDA has been tested on Linux and Windows binaries taken from the SPEC CPU2017, SPEC CPU2006 benchmark suits, and the BAP corpus on two popular disassembly tasks, recovering function boundaries and instructions. XDA achieves $17.2\%$ higher accuracy 
than the second-best tool in the function boundary task and $99.7\% $ F1 score at recovering assembly instructions.

\subsection{Attack management and threat intelligence}\label{HMM}

Recently, there has been a noticeable increase in complex attacks that involve multiple attack phases. These sophisticated attacks are designed to exploit vulnerabilities at different stages of a system or network, often evading traditional security defenses. Detecting and mitigating such attacks with ML-based techniques requires the models to be able of analyzing diverse indicators of compromise across different stages and ultimately reveal the motives of the attacker. The Hidden Markov Model (HMM) is a popular machine learning technique widely used to address sequential attacks~\cite{chadza2020learning}. HMMs are probabilistic models that utilize state transitions and emission probability distributions to recognize and model different stages or states within a system or network.

 Learning parameters in HMMs can be challenging, especially in the security domains where labeled datasets may be unavailable or limited. This issue is common to security due to the evolving nature of attacks, the scarcity of labeled attack data, and the need for up-to-date training data. To address these challenges,  the work in~\cite{chadza2020learning} assesses the effectiveness of transfer learning techniques for sequential network attacks. The primary driver of this work is to leverage prior modelling capability and knowledge derived from known, labelled datasets and use it to efficiently develop models for new, yet unlabelled, datasets in an unsupervised fashion. They combined transfer learning with HMMs where the parameters of the source's HMM model is used as the starting point for the target's HMM model. They evaluated several HMM techniques for detecting the attack stages in the network traffic, as well as, forecasting the next most probable attack stage. 

\begin{figure*}[t]
  \centering
  \includegraphics[width=\linewidth]{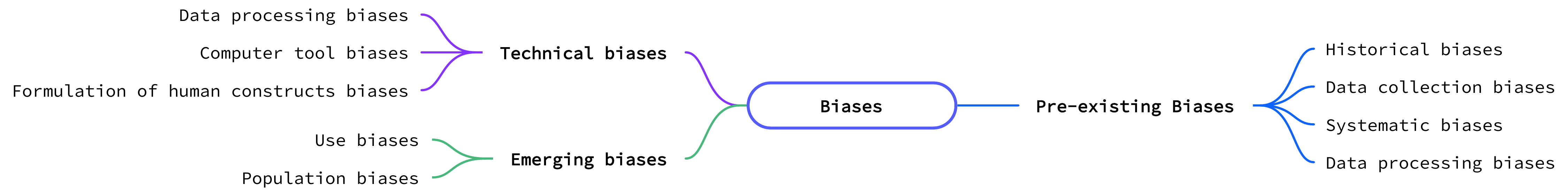}
  \caption{A taxonomy of the currently existing forms of bias (based on Figure 1 from~\cite{richardson2021framework}). }
  \label{ethics}
\end{figure*}

\subsection{TL for LLM in the context of security}\label{forensic}



The combination of transfer learning with Natural Language Processing (NLP), Large Language Models (LLM) presents a compelling solution to assist forensic analysis and vulnerability assessment. Models such as BERT, adept at absorbing intricate semantic and contextual nuances from extensive text datasets, can be tailored to suit the unique requirements of forensic analysis and vulnerability assessment. Within forensic analysis, NLP tasks like question answering~\cite{mccann2018natural} and language modeling~\cite{radford2018improving} have great potential. For instance, NLP models proficient in comprehending textual queries can significantly improve investigations by extracting relevant information and providing insights. Similarly, in vulnerability assessment, transfer learning and NLP techniques can be leveraged to analyze textual data from security reports and identify potential weaknesses or threats. By harnessing sentiment analysis techniques, security professionals can gain deeper insights into the severity and implications of identified vulnerabilities. In summary, the integration of transfer learning with NLP not only enhances the efficiency of forensic analysis and vulnerability assessment but also enables a more comprehensive understanding of textual data, empowering cybersecurity professionals to make informed decisions and mitigate risks effectively.

%% file: sections/4_challenges.tex
\section{Challenges and Considerations}\label{problems}


Despite the advancements made in transfer learning for security, there are several open problems that need to be addressed. These problems can be categorized as either common challenges encountered in computer vision tasks or specific issues that are unique to security applications.

\paragraph{\textbf{Discrepancy between source and target domains}} Addressing domain discrepancy is crucial because the pre-trained models might not generalize well to the target domain due to the differences in data distribution. The features learned in the source domain may not be directly applicable or transferable to the target domain, leading to a decrease in performance~\cite{li2023maximal}.  For example, source domain data might originate from traditional networks, while target data could be derived from IoT networks. Consequently, due to the distribution shift in the collected features, the source data may not necessarily prove beneficial for the target network. To overcome this challenge, researchers have been exploring more sophisticated techniques such as domain adaptation methods~\cite{li2023maximal, ganin2016domain, singla2020preparing}. 
These methods aim to bridge the gap between the source and target domains by aligning the feature representations. 

While many recent approaches for transfer learning focus on the single source domain setting, there is potential value in utilizing data from multiple related domains. In the context of security applications, it is possible to have access to several labeled datasets collected over time or from different vendors, which can serve as source domains for transfer learning. The key research challenge lies in assessing the suitability of a source domain for knowledge transfer and determining the optimal amount of knowledge to be transferred, with a general principle of transferring more knowledge for closely related domains and less knowledge for dissimilar domains. In 
Section~\ref{directions}, we will delve deeper into this topic and explore different strategies and techniques for multi-source domain adaptation.

\paragraph{\textbf{Imbalanced data}} A common limitation in many transfer learning approaches is the assumption of a balanced target domain dataset, even when the target domain has limited labels. However, real-life security datasets often exhibit imbalanced class distributions, which can hurt the performance of DA models.
Several approaches can be used to address class imbalance, including weighted loss functions, and the oversampling or undersampling of training data in the minority and majority classes, respectively. Nevertheless, the effectiveness of these methods heavily relies on the characteristics of the datasets and the specific learning task at hand. In Section~\ref{directions}, we discuss research directions related to the  combination of transfer learning techniques with generative models to address the imbalanced data issue.

\paragraph{\textbf{New attack labels}}
Existing research on applying DA techniques in intrusion detection has shown significant progress in recent years~\cite{singla2020preparing, lin2018idsgan, zhao2019transfer}. However, these methods have primarily been evaluated under a closed set setting, where both the source and target domains consist of exactly the same classes.

Earlier Panareda et al.~\cite{panareda2017open} had introduced the concept of open set scenarios, which represents a more realistic scenario. In open set scenarios, only a few labels are shared between the source and target data. This is particularly relevant in security applications, where the source and target data may have limited or no overlapping attack labels, and the target data might introduce new attack labels not present in the source domain. Surprisingly, to the best of our knowledge, no previous work has proposed a transfer learning method suitable for both closed and open set scenarios in the context of security applications.

\paragraph{\textbf{Adversarial robustness}}
Robustness refers to the ability of a trained model to do well when there are changes in the environment where the model is deployed. There are various reasons for these changes, including malicious attacks, unmodeled phenomena, undetected biases, or significant changes in data. 

Transfer learning techniques exhibit greater generality compared to models based solely on DL because transfer learning leverages knowledge acquired from multiple domains. 
However, transfer learning techniques which use neural networks could still be vulnerable to adversarial attacks. Adversarial attacks involve malicious actors manipulating input data in a way that can bypass the defense mechanisms of models, including intrusion detection systems. Attackers exploit vulnerabilities in the models' decision boundaries, making subtle changes to input data that may go unnoticed by humans but can mislead the transfer learning model into making incorrect predictions. These adversarial attacks pose a significant challenge. As transfer learning models are often employed in critical security tasks, such as identifying network intrusions, the potential for malicious actors to manipulate data and deceive the models raises serious concerns about the reliability and trustworthiness of these systems.

\paragraph{\textbf{Confirmation bias}} Confirmation bias affects machine learning in general and leads to degraded performance. In application to security, models trained with missing or biased data can lead to incorrect inferences and classification of vulnerabilities, threats, malware and so on~\cite{lemay2018cognitive, krawczyk2013measuring}. Domain adaptation and transfer learning can alleviate the existing confirmation biases with  targeted datasets. Confirmation bias maybe injected by transfer learning as well. That can lead to the customized model to be biased in its security specific decisions. 

\paragraph{\textbf{Ethical risks and fairness issues}}  Transfer learning can be a powerful tool for security applications. However, when the models or datasets used are biased, ethical implications arise that can potentially limit their effectiveness and fairness. A taxonomy of type of bias is examined in~\cite{richardson2021framework} and  shown in Figure~\ref{ethics}.  Pre-existing biases are connected to individuals or institutions and occur when human inclinations or societal stereotypes influence the data or the model, as described by~\cite{richardson2021framework}. This type of bias subsequently becomes evident in data gathered over a period of time. Technical biases, on the other hand, arise due to the limitations inherent in computer and data technology. This might involve how the choice of attributes, models, or training methodologies could instigate biases that are not directly tied to the entity executing the training, but rather to the shortcomings of those procedures~\cite{richardson2021framework}. Lastly, emerging biases are those that appear following the implementation of a model. This category of bias can manifest in two forms: population bias and use bias. Population bias originates from the model's inability to adequately represent its demographic post-deployment. Use biases correspond to the biases that people develop after engaging with a model ~\cite{richardson2021framework}. 

Numerous examples highlight the ways in which data and model biases can compromise fairness. This issue becomes particularly evident in transfer learning. If the originating model is trained using biased data, this bias is likely to be propagated to the target model. Moreover, if the data from the source does not sufficiently reflect the diversity of the population for which the resulting model is intended, it could induce algorithmic bias. This could consequently lead to the unfair treatment of groups that are underrepresented in the data.

\paragraph{\textbf{Data privacy}} 
When it comes to security, it is crucial to utilize machine learning models that prioritize privacy when dealing with security tasks. On the other hand, transfer learning approaches require access to the actual source dataset, as opposed to using a DL model trained on the source data. This can pose a significant obstacle if the organization that owns the dataset is unwilling to share the data that might have sensitive information, such as IP addresses. As a result, the potential for collaborative efforts aimed at creating effective network intrusion detection models capable of addressing different network types and emerging attacks is limited. Simply anonymizing the dataset is not sufficient, as one can easily leverage additional information from related data sources~\cite{singla2022dp}.

%% file: sections/5_directions.tex
\section{Research Directions}\label{directions}

\paragraph{\textbf{Dealing with imbalanced class distribution in security datasets}}
Recent techniques leverage generative models, such as generative adversarial networks (GAN)~\cite{goodfellow2020generative}, autoencoders (AE)~\cite{engel2017neural} and diffusion models~\cite{ho2022cascaded}, to produce synthetic images for augementing imbalanced data. Generative models are able to generate realistic data samples, making them poised to have a significant influence in the coming years.

Synthetic data offers a convenient and cost-effective alternative to acquiring real-world data.  However,  systems built using such data often encounter failures when deployed in real-world scenarios. This issue stems from the distribution disparity between synthetic and real data, commonly referred to as the sim-to-real problem~\cite{peng2018sim}.

\begin{figure}[t]
  \centering
  \includegraphics[width=\linewidth]{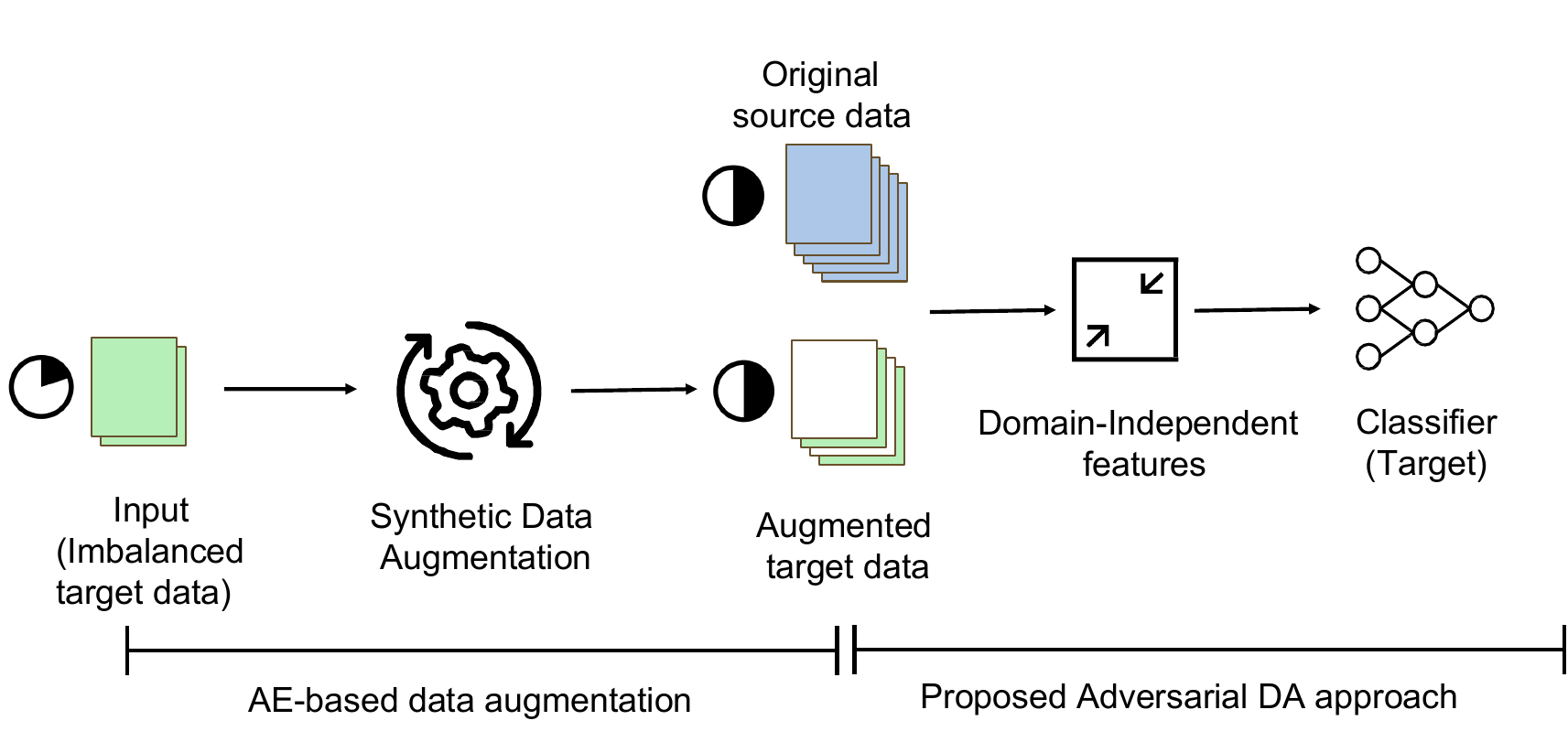}
  \caption{Combining transfer learning with the autoencoder-based data augmentation.}
  \label{AE}
\end{figure}

Li et al.~\cite{li2023building} have recentlyd propose a pipeline that addresses these limitations. The pipeline integrates two key components
(see Figure~\ref{AE}). The first component is an autoencoder-based method that enhances the target data by generating synthetic data for underrepresented classes using Gaussian noise and the encoder's learned latent space. This approach effectively tackles the issue of imbalanced data.  The second component introduces a novel transfer learning domain architecture based on adversarial DA. This architecture is designed to handle challenges associated with limited training data and the discrepancy between synthetic and real data distributions. By leveraging adversarial techniques, the pipeline aims to bridge the gap between the synthetic and real domains.  However, the evaluation of this approach has focused solely on image data, raising the question of whether this approach can be effectively applied to security datasets. 

When selecting a generative model for a particular task, it is crucial to consider the advantages, limitations, and costs associated with each model~\cite{oussidi2018deep}. While GANs are capable of generating high-quality image data, they are known to suffer from training instability and can be susceptible to mode collapse during the training process~\cite{salimans2016improved}.

More recently, diffusion models have gained significant attention due to their impressive generation capabilities~\cite{croitoru2022diffusion, ho2022cascaded}. However, diffusion models typically come with high computational costs due to the iterative steps involved in their training process~\cite{croitoru2022diffusion}. As a result, they may not be suitable for tasks that require time-sensitive operations.

 It is important to assess the performance, efficiency, and suitability of different generative models in the context of security applications. Further investigation and experimentation are necessary to determine which generative model works best for security datasets.

\paragraph{\textbf{Privacy preserving TL}}
 Several approaches have been proposed for training DL models with Differential Privacy (DP) guarantees. DP also provides a means to measure privacy loss, which is quantified as the privacy budget. A lower privacy budget indicates a higher level of privacy protection, resulting in relatively stronger privacy guarantees for individuals. 
 
 Singla et al.~\cite{singla2022dp} have recently proposed a differentially-private adversarial DA workflow (DP-ADA) workflow that ensures the privacy of the source dataset while enabling adversarial domain adaptation. The process involves several steps: \ding{182} Organization A possesses a NIDS dataset, which includes labeled packet captures of benign and attack traffic from their network. This dataset is considered the source dataset.  \ding {183} Organization A applies the DP-CGAN approach to train on the source dataset, which can generate a synthetic dataset that closely resembles the real source data with DP guarantees.  \ding{184} Organization A shares the trained model only with Organization B. \ding{185} Organization B uses the model to generate a synthetic dataset that mimics the source dataset. \ding{186} Orgnization B performs adversarial DA  by employing a small labeled target dataset created using network traffic from their network. The result is a highly accurate NID classifier that operates on both the target network and the source dataset. Evaluating this workflow for other security domains like malware and botnet detection would be interesting. This evaluation would provide insights into the effectiveness and applicability of the DP-ADA workflow across various security contexts.

\paragraph{\textbf{Multi-source approach in security related tasks}}

In recent years, DA has received significant attention in research. However, the majority of theoretical results and algorithms have primarily focused on the single-source-single-target adaptation setting. In real-world application scenarios, it is common to encounter situations where labeled data is available from multiple domains, each with its own distinct distribution. This scenario, known as multi-source domain adaptation, poses unique challenges that have not yet been extensively explored in security-related tasks. However, multi-source DA has shown promising effectiveness in other domains such as natural language processing and computer vision tasks. In the following paragraph, we briefly describe a multi-source DA approach called MDAN, which is an extension of  the single-source DA approach shown in Figure~\ref{GAN}.

Zhao et al.~\cite{zhao2018multiple} have proved a new generalization bound for multi-source DA when there are multiple source domains with labeled data and one target domain with unlabeled data.  The optimization problem for each model is a minimax saddle point problem, which can be interpreted as a minimax game with two participants competing against each other to learn invariant features. The feature extraction, domain classification, and task learning are combined in one training process~\cite{zhao2018multiple}. 

While MDAN demonstrates superior performance when compared to other competing methods on three real-world datasets, encompassing a sentiment analysis task, a digit classification task, and a visual vehicle counting task, the question of its effectiveness in security-related tasks remains an open area of investigation. Additionally, it's important to note that MDAN is limited to closed-set scenarios, where both the source and target domains have the same classes. 

Although the majority of research efforts are directed toward leveraging data collected from diverse sources, an intriguing and relatively uncharted domain lies in a hybrid setting. In this setting, some sources share their models while others share their datasets. The adaptability of this approach necessitates the integration of transfer learning with other machine learning techniques, paving the way for innovative solutions in various security tasks.

\paragraph{\textbf{Integration of TL with FL: applications in security}}

One of the primary benefits of federated learning (FL) is the enhanced protection of user data. Data remains on the local device during training, which can reduce the risk of data leakage. Moreover, By enabling local model training, FL utilizes computational capabilities of edge devices, such as smartphones or Internet of Things (IoT) devices, reducing dependency on centralized servers.

Previous methods generally assume that data is centralized on a single server. This assumption restricts their utility in a distributed learning environment. The federated setup introduces a variety of new challenges. First of all, the data are stored locally and cannot be shared in FL, while conventional domain adaptation techniques require access to both the source data  and target data.  Additionally, the knowledge extracted from the source nodes tends to be intertwined, which  could potentially lead to a phenomenon known as negative transfer. This is when the performance of the target node deteriorates due to the irrelevant knowledge from the source nodes.

To solve the above problems,  a solution called Federated Adversarial Domain Adaptation (FADA) was proposed in~\cite{peng2019federated}. Within this method, models are trained separately on each source domain, then their gradients are assembled via a dynamic attention mechanism, which serves to update the target model. Once again, FADA is only tested on vision and linguistic
benchmarks, and its effectiveness for the security task remains to be investigated.


\paragraph{\textbf{Integration of TL with RL: applications in security}}
RL techniques are designed to identify the most optimal strategy for solving sequential decision-making problems, aiming to maximize long-term objectives within such scenarios. These algorithms achieve this by learning from past experiences, reinforcing ``good" behaviors, and avoiding ``bad" ones. Consequently, RL techniques are particularly well-suited for security applications that involve sequential decision-making. Combining transfer learning with reinforcement learning can result in security solutions that are not only adaptive but also highly effective in dynamic environments. For example, transfer learning methods can be leveraged to expedite the process of RL-based security policy optimization~\cite{zhang2021reinforcement}. Defining performance metrics for security application is often challenging, as optimization objectives typically involves multiple functionalities, requiring careful design.

%% file: sections/6_conclusion.tex
\section{Conclusion}\label{conclusion}

The increasing reliance on digital systems and information makes cybersecurity more critical than ever. With considerable advancements in complexity, efficiency, and applicability, machine learning presents a vast opportunity for bolstering cybersecurity measures. As a result, a wide array of security tasks, notably policy training, anomaly detection, and electronic forensics, have substantially employed machine learning methodologies.

However, despite the very encouraging state of the art in applying machine learning to security functions, the recent success has been partially attributed to the large-scale datasets on which they are trained. With the maturity of today’s neural network architectures, for many practical applications, the bottleneck will be whether we can efficiently get the data we need to develop systems that work well. While addressing data scarcity through transfer learning is not new in fields like computer vision and natural language processing, its application in the security domain is still in its early stages. In this paper, we have comprehensively reviewed the current research literature concerning the application of transfer learning to various security functions. We have identified and discussed the key challenges that need to be acknowledged and addressed when employing transfer learning techniques in security tasks. Furthermore, we have outlined some potential future research directions in transfer learning. We believe that transfer learning techniques will play an important role in the field of security.\\

\noindent
{\bf Acknowledgements.} This work has been partially supported by NSF grant  IIS 2229876.

%% file: main.bbl
\begin{thebibliography}{49}
\providecommand{\natexlab}[1]{#1}
\providecommand{\url}[1]{\texttt{#1}}
\expandafter\ifx\csname urlstyle\endcsname\relax
  \providecommand{\doi}[1]{doi: #1}\else
  \providecommand{\doi}{doi: \begingroup \urlstyle{rm}\Url}\fi

\bibitem[Abu~Jabal et~al.(2020)Abu~Jabal, Bertino, Lobo, Law, Russo, Calo, and
  Verma]{abu2020polisma}
Amani Abu~Jabal, Elisa Bertino, Jorge Lobo, Mark Law, Alessandra Russo,
  Seraphin Calo, and Dinesh Verma.
\newblock Polisma-a framework for learning attribute-based access control
  policies.
\newblock In \emph{Computer Security--ESORICS 2020: 25th European Symposium on
  Research in Computer Security, ESORICS 2020, Guildford, UK, September 14--18,
  2020, Proceedings, Part I 25}, pages 523--544. Springer, 2020.

\bibitem[Abu~Jabal et~al.(2023)Abu~Jabal, Bertino, Lobo, Verma, Calo, and
  Russo]{abu2023flap}
Amani Abu~Jabal, Elisa Bertino, Jorge Lobo, Dinesh Verma, Seraphin Calo, and
  Alessandra Russo.
\newblock Flap-a federated learning framework for attribute-based access
  control policies.
\newblock In \emph{Proceedings of the Thirteenth ACM Conference on Data and
  Application Security and Privacy}, pages 263--272, 2023.

\bibitem[Bertino et~al.(2023)Bertino, Bhardwaj, Cicala, Gong, Karim, Katsis,
  Lee, Li, and Mahgoub]{bertino2023machine}
Elisa Bertino, Sonam Bhardwaj, Fabrizio Cicala, Sishuai Gong, Imtiaz Karim,
  Charalampos Katsis, Hyunwoo Lee, Adrian~Shuai Li, and Ashraf~Y Mahgoub.
\newblock \emph{Machine Learning Techniques for Cybersecurity}.
\newblock Springer Nature, 2023.

\bibitem[Chadza et~al.(2020)Chadza, Kyriakopoulos, and
  Lambotharan]{chadza2020learning}
Timothy Chadza, Konstantinos~G Kyriakopoulos, and Sangarapillai Lambotharan.
\newblock Learning to learn sequential network attacks using hidden markov
  models.
\newblock \emph{IEEE Access}, 8:\penalty0 134480--134497, 2020.

\bibitem[Croitoru et~al.(2022)Croitoru, Hondru, Ionescu, and
  Shah]{croitoru2022diffusion}
Florinel-Alin Croitoru, Vlad Hondru, Radu~Tudor Ionescu, and Mubarak Shah.
\newblock Diffusion models in vision: A survey.
\newblock \emph{arXiv preprint arXiv:2209.04747}, 2022.

\bibitem[Deng et~al.(2009)Deng, Dong, Socher, Li, Li, and
  Fei-Fei]{deng2009imagenet}
Jia Deng, Wei Dong, Richard Socher, Li-Jia Li, Kai Li, and Li~Fei-Fei.
\newblock Imagenet: A large-scale hierarchical image database.
\newblock In \emph{2009 IEEE conference on computer vision and pattern
  recognition}, pages 248--255. Ieee, 2009.

\bibitem[Engel et~al.(2017)Engel, Resnick, Roberts, Dieleman, Norouzi, Eck, and
  Simonyan]{engel2017neural}
Jesse Engel, Cinjon Resnick, Adam Roberts, Sander Dieleman, Mohammad Norouzi,
  Douglas Eck, and Karen Simonyan.
\newblock Neural audio synthesis of musical notes with wavenet autoencoders.
\newblock In \emph{International Conference on Machine Learning}, pages
  1068--1077. PMLR, 2017.

\bibitem[Ganin et~al.(2016)Ganin, Ustinova, Ajakan, Germain, Larochelle,
  Laviolette, Marchand, and Lempitsky]{ganin2016domain}
Yaroslav Ganin, Evgeniya Ustinova, Hana Ajakan, Pascal Germain, Hugo
  Larochelle, Fran{\c{c}}ois Laviolette, Mario Marchand, and Victor Lempitsky.
\newblock Domain-adversarial training of neural networks.
\newblock \emph{The journal of machine learning research}, 17\penalty0
  (1):\penalty0 2096--2030, 2016.

\bibitem[Goodfellow et~al.(2020)Goodfellow, Pouget-Abadie, Mirza, Xu,
  Warde-Farley, Ozair, Courville, and Bengio]{goodfellow2020generative}
Ian Goodfellow, Jean Pouget-Abadie, Mehdi Mirza, Bing Xu, David Warde-Farley,
  Sherjil Ozair, Aaron Courville, and Yoshua Bengio.
\newblock Generative adversarial networks.
\newblock \emph{Communications of the ACM}, 63\penalty0 (11):\penalty0
  139--144, 2020.

\bibitem[Ho et~al.(2022)Ho, Saharia, Chan, Fleet, Norouzi, and
  Salimans]{ho2022cascaded}
Jonathan Ho, Chitwan Saharia, William Chan, David~J Fleet, Mohammad Norouzi,
  and Tim Salimans.
\newblock Cascaded diffusion models for high fidelity image generation.
\newblock \emph{J. Mach. Learn. Res.}, 23\penalty0 (47):\penalty0 1--33, 2022.

\bibitem[Krawczyk et~al.(2013)Krawczyk, Bartlett, Kantarcioglu, Hamlen, and
  Thuraisingham]{krawczyk2013measuring}
Daniel Krawczyk, James Bartlett, Murat Kantarcioglu, Kevin Hamlen, and Bhavani
  Thuraisingham.
\newblock Measuring expertise and bias in cyber security using cognitive and
  neuroscience approaches.
\newblock In \emph{2013 IEEE International Conference on Intelligence and
  Security Informatics}, pages 364--367. IEEE, 2013.

\bibitem[Kumar and Janet(2022)]{kumar2022dtmic}
Sanjeev Kumar and B~Janet.
\newblock Dtmic: Deep transfer learning for malware image classification.
\newblock \emph{Journal of Information Security and Applications}, 64:\penalty0
  103063, 2022.

\bibitem[Kumar et~al.(2021)]{kumar2021mcft}
Sushil Kumar et~al.
\newblock Mcft-cnn: Malware classification with fine-tune convolution neural
  networks using traditional and transfer learning in internet of things.
\newblock \emph{Future Generation Computer Systems}, 125:\penalty0 334--351,
  2021.

\bibitem[Lemay and Leblanc(2018)]{lemay2018cognitive}
Antoine Lemay and Sylvain Leblanc.
\newblock Cognitive biases in cyber decision-making.
\newblock In \emph{Proceedings of the 13th International Conference on Cyber
  Warfare and Security}, page 395, 2018.

\bibitem[Li et~al.(2023{\natexlab{a}})Li, Bertino, Dang, Singla, Tu, and
  Wegman]{li2023maximal}
Adrian~Shuai Li, Elisa Bertino, Xuan-Hong Dang, Ankush Singla, Yuhai Tu, and
  Mark~N Wegman.
\newblock Maximal domain independent representations improve transfer learning.
\newblock \emph{arXiv preprint arXiv:2306.00262}, 2023{\natexlab{a}}.

\bibitem[Li et~al.(2023{\natexlab{b}})Li, Bertino, Wu, and Wu]{li2023building}
Adrian~Shuai Li, Elisa Bertino, Rih-Teng Wu, and Ting-Yan Wu.
\newblock Building manufacturing deep learning models with minimal and
  imbalanced training data using domain adaptation and data augmentation.
\newblock In \emph{2023 IEEE International Conference on Industrial Technology
  (ICIT)}, pages 1--8. IEEE, 2023{\natexlab{b}}.

\bibitem[Li et~al.(2019)Li, Zhang, Ding, and Li]{li2019diagnosing}
Xiang Li, Wei Zhang, Qian Ding, and Xu~Li.
\newblock Diagnosing rotating machines with weakly supervised data using deep
  transfer learning.
\newblock \emph{IEEE transactions on industrial informatics}, 16\penalty0
  (3):\penalty0 1688--1697, 2019.

\bibitem[Li et~al.(2012)Li, Xia, Zhang, Yan, Ai, and Dai]{li2012efficient}
Yinhui Li, Jingbo Xia, Silan Zhang, Jiakai Yan, Xiaochuan Ai, and Kuobin Dai.
\newblock An efficient intrusion detection system based on support vector
  machines and gradually feature removal method.
\newblock \emph{Expert systems with applications}, 39\penalty0 (1):\penalty0
  424--430, 2012.

\bibitem[Lin et~al.(2015)Lin, Ke, and Tsai]{lin2015cann}
Wei-Chao Lin, Shih-Wen Ke, and Chih-Fong Tsai.
\newblock Cann: An intrusion detection system based on combining cluster
  centers and nearest neighbors.
\newblock \emph{Knowledge-based systems}, 78:\penalty0 13--21, 2015.

\bibitem[Lin et~al.(2018)Lin, Shi, and Xue]{lin2018idsgan}
Zilong Lin, Yong Shi, and Zhi Xue.
\newblock Idsgan: Generative adversarial networks for attack generation against
  intrusion detection.
\newblock \emph{arXiv preprint arXiv:1809.02077}, 2018.

\bibitem[Lyu et~al.(2022)Lyu, Fang, Zhang, Sun, Ma, Bertino, Lu, and
  Li]{lyu2022goshawk}
Yunlong Lyu, Yi~Fang, Yiwei Zhang, Qibin Sun, Siqi Ma, Elisa Bertino, Kangjie
  Lu, and Juanru Li.
\newblock Goshawk: Hunting memory corruptions via structure-aware and
  object-centric memory operation synopsis.
\newblock In \emph{2022 IEEE Symposium on Security and Privacy (SP)}, pages
  2096--2113. IEEE, 2022.

\bibitem[McCann et~al.(2018)McCann, Keskar, Xiong, and
  Socher]{mccann2018natural}
Bryan McCann, Nitish~Shirish Keskar, Caiming Xiong, and Richard Socher.
\newblock The natural language decathlon: Multitask learning as question
  answering.
\newblock \emph{arXiv preprint arXiv:1806.08730}, 2018.

\bibitem[Mirsky et~al.(2018)Mirsky, Doitshman, Elovici, and
  Shabtai]{mirsky2018kitsune}
Yisroel Mirsky, Tomer Doitshman, Yuval Elovici, and Asaf Shabtai.
\newblock Kitsune: an ensemble of autoencoders for online network intrusion
  detection.
\newblock \emph{arXiv preprint arXiv:1802.09089}, 2018.

\bibitem[Mudgerikar et~al.(2019)Mudgerikar, Sharma, and
  Bertino]{mudgerikar2019spion}
Anand Mudgerikar, Puneet Sharma, and Elisa Bertino.
\newblock E-spion: A system-level intrusion detection system for iot devices.
\newblock In \emph{Proceedings of the 2019 ACM Asia conference on computer and
  communications security}, pages 493--500, 2019.

\bibitem[Oussidi and Elhassouny(2018)]{oussidi2018deep}
Achraf Oussidi and Azeddine Elhassouny.
\newblock Deep generative models: Survey.
\newblock In \emph{2018 International conference on intelligent systems and
  computer vision (ISCV)}, pages 1--8. IEEE, 2018.

\bibitem[Pan and Yang(2009)]{pan2009survey}
Sinno~Jialin Pan and Qiang Yang.
\newblock A survey on transfer learning.
\newblock \emph{IEEE Transactions on knowledge and data engineering},
  22\penalty0 (10):\penalty0 1345--1359, 2009.

\bibitem[Panareda~Busto and Gall(2017)]{panareda2017open}
Pau Panareda~Busto and Juergen Gall.
\newblock Open set domain adaptation.
\newblock In \emph{Proceedings of the IEEE international conference on computer
  vision}, pages 754--763, 2017.

\bibitem[Pei et~al.(2020)Pei, Guan, Williams-King, Yang, and Jana]{pei2020xda}
Kexin Pei, Jonas Guan, David Williams-King, Junfeng Yang, and Suman Jana.
\newblock Xda: Accurate, robust disassembly with transfer learning.
\newblock \emph{arXiv preprint arXiv:2010.00770}, 2020.

\bibitem[Peng et~al.(2019)Peng, Huang, Zhu, and Saenko]{peng2019federated}
Xingchao Peng, Zijun Huang, Yizhe Zhu, and Kate Saenko.
\newblock Federated adversarial domain adaptation.
\newblock \emph{arXiv preprint arXiv:1911.02054}, 2019.

\bibitem[Peng et~al.(2018)Peng, Andrychowicz, Zaremba, and Abbeel]{peng2018sim}
Xue~Bin Peng, Marcin Andrychowicz, Wojciech Zaremba, and Pieter Abbeel.
\newblock Sim-to-real transfer of robotic control with dynamics randomization.
\newblock In \emph{IEEE international conference on robotics and automation
  (ICRA)}, pages 3803--3810. IEEE, 2018.

\bibitem[Radford et~al.(2018)Radford, Narasimhan, Salimans, Sutskever,
  et~al.]{radford2018improving}
Alec Radford, Karthik Narasimhan, Tim Salimans, Ilya Sutskever, et~al.
\newblock Improving language understanding by generative pre-training.
\newblock 2018.

\bibitem[Rajpal et~al.(2017)Rajpal, Blum, and Singh]{rajpal2017not}
Mohit Rajpal, William Blum, and Rishabh Singh.
\newblock Not all bytes are equal: Neural byte sieve for fuzzing.
\newblock \emph{arXiv preprint arXiv:1711.04596}, 2017.

\bibitem[Richardson and Gilbert(2021)]{richardson2021framework}
Brianna Richardson and Juan~E Gilbert.
\newblock A framework for fairness: A systematic review of existing fair ai
  solutions.
\newblock \emph{arXiv preprint arXiv:2112.05700}, 2021.

\bibitem[Rustam et~al.(2023)Rustam, Ashraf, Jurcut, Bashir, and
  Zikria]{rustam2023malware}
Furqan Rustam, Imran Ashraf, Anca~Delia Jurcut, Ali~Kashif Bashir, and
  Yousaf~Bin Zikria.
\newblock Malware detection using image representation of malware data and
  transfer learning.
\newblock \emph{Journal of Parallel and Distributed Computing}, 172:\penalty0
  32--50, 2023.

\bibitem[Rutkowski et~al.(2013)Rutkowski, Jaworski, Pietruczuk, and
  Duda]{rutkowski2013decision}
Leszek Rutkowski, Maciej Jaworski, Lena Pietruczuk, and Piotr Duda.
\newblock Decision trees for mining data streams based on the gaussian
  approximation.
\newblock \emph{IEEE Transactions on Knowledge and Data Engineering},
  26\penalty0 (1):\penalty0 108--119, 2013.

\bibitem[Salimans et~al.(2016)Salimans, Goodfellow, Zaremba, Cheung, Radford,
  and Chen]{salimans2016improved}
Tim Salimans, Ian Goodfellow, Wojciech Zaremba, Vicki Cheung, Alec Radford, and
  Xi~Chen.
\newblock Improved techniques for training gans.
\newblock \emph{Advances in neural information processing systems}, 29, 2016.

\bibitem[Shao et~al.(2014)Shao, Zhu, and Li]{shao2014transfer}
Ling Shao, Fan Zhu, and Xuelong Li.
\newblock Transfer learning for visual categorization: A survey.
\newblock \emph{IEEE transactions on neural networks and learning systems},
  26\penalty0 (5):\penalty0 1019--1034, 2014.

\bibitem[Shao et~al.(2018)Shao, McAleer, Yan, and Baldi]{shao2018highly}
Siyu Shao, Stephen McAleer, Ruqiang Yan, and Pierre Baldi.
\newblock Highly accurate machine fault diagnosis using deep transfer learning.
\newblock \emph{IEEE Transactions on Industrial Informatics}, 15\penalty0
  (4):\penalty0 2446--2455, 2018.

\bibitem[Singla and Bertino(2022)]{singla2022dp}
Ankush Singla and Elisa Bertino.
\newblock Dp-ada: Differentially private adversarial domain adaptation for
  training deep learning based network intrusion detection systems.
\newblock In \emph{2022 IEEE 8th International Conference on Collaboration and
  Internet Computing (CIC)}, pages 89--98. IEEE, 2022.

\bibitem[Singla et~al.(2019)Singla, Bertino, and Verma]{singla2019overcoming}
Ankush Singla, Elisa Bertino, and Dinesh Verma.
\newblock Overcoming the lack of labeled data: Training intrusion detection
  models using transfer learning.
\newblock In \emph{2019 IEEE International Conference on Smart Computing
  (SMARTCOMP)}, pages 69--74. IEEE, 2019.

\bibitem[Singla et~al.(2020)Singla, Bertino, and Verma]{singla2020preparing}
Ankush Singla, Elisa Bertino, and Dinesh Verma.
\newblock Preparing network intrusion detection deep learning models with
  minimal data using adversarial domain adaptation.
\newblock In \emph{Proceedings of the 15th ACM Asia conference on computer and
  communications security}, pages 127--140, 2020.

\bibitem[Sun et~al.(2018)Sun, Ma, Zhao, Tian, Yan, and Chen]{sun2018deep}
Chuang Sun, Meng Ma, Zhibin Zhao, Shaohua Tian, Ruqiang Yan, and Xuefeng Chen.
\newblock Deep transfer learning based on sparse autoencoder for remaining
  useful life prediction of tool in manufacturing.
\newblock \emph{IEEE transactions on industrial informatics}, 15\penalty0
  (4):\penalty0 2416--2425, 2018.

\bibitem[Tang et~al.(2020)Tang, Yang, Li, Meng, Wang, Li, Sun, Pei, Wei, Xu,
  et~al.]{tang2020zerowall}
Ruming Tang, Zheng Yang, Zeyan Li, Weibin Meng, Haixin Wang, Qi~Li, Yongqian
  Sun, Dan Pei, Tao Wei, Yanfei Xu, et~al.
\newblock Zerowall: Detecting zero-day web attacks through encoder-decoder
  recurrent neural networks.
\newblock In \emph{IEEE INFOCOM 2020-IEEE Conference on Computer
  Communications}, pages 2479--2488. IEEE, 2020.

\bibitem[Van~Wyk et~al.(2019)Van~Wyk, Wang, Khojandi, and Masoud]{van2019real}
Franco Van~Wyk, Yiyang Wang, Anahita Khojandi, and Neda Masoud.
\newblock Real-time sensor anomaly detection and identification in automated
  vehicles.
\newblock \emph{IEEE Transactions on Intelligent Transportation Systems},
  21\penalty0 (3):\penalty0 1264--1276, 2019.

\bibitem[Wang et~al.(2018)Wang, Shi, Li, Chen, Ding, and
  Duan]{wang2018distributed}
Jingyu Wang, Dongyuan Shi, Yinhong Li, Jinfu Chen, Hongfa Ding, and Xianzhong
  Duan.
\newblock Distributed framework for detecting pmu data manipulation attacks
  with deep autoencoders.
\newblock \emph{IEEE Transactions on smart grid}, 10\penalty0 (4):\penalty0
  4401--4410, 2018.

\bibitem[Wang and Deng(2018)]{wang2018deep}
Mei Wang and Weihong Deng.
\newblock Deep visual domain adaptation: A survey.
\newblock \emph{Neurocomputing}, 312:\penalty0 135--153, 2018.

\bibitem[Zhang et~al.(2021)Zhang, Mudgerikar, Singla, Leung, Bertino, Verma,
  Chan, Melrose, and Tucker]{zhang2021reinforcement}
Ziyao Zhang, Anand Mudgerikar, Ankush Singla, Kin~K Leung, Elisa Bertino,
  Dinesh Verma, Kevin Chan, John Melrose, and Jeremy Tucker.
\newblock Reinforcement and transfer learning for distributed analytics in
  fragmented software defined coalitions.
\newblock In \emph{Artificial Intelligence and Machine Learning for
  Multi-Domain Operations Applications III}, volume 11746, pages 442--452.
  SPIE, 2021.

\bibitem[Zhao et~al.(2018)Zhao, Zhang, Wu, Gordon, et~al.]{zhao2018multiple}
Han Zhao, Shanghang Zhang, Guanhang Wu, Geoffrey~J Gordon, et~al.
\newblock Multiple source domain adaptation with adversarial learning.
\newblock 2018.

\bibitem[Zhao et~al.(2019)Zhao, Shetty, Pan, Kamhoua, and
  Kwiat]{zhao2019transfer}
Juan Zhao, Sachin Shetty, Jan~Wei Pan, Charles Kamhoua, and Kevin Kwiat.
\newblock Transfer learning for detecting unknown network attacks.
\newblock \emph{EURASIP Journal on Information Security}, 2019:\penalty0 1--13,
  2019.

\end{thebibliography}
